# Synthesis and Laser Processing of ZnO Nanocrystalline Thin Films

I. Ozerov[1], D. Nelson[2], A.V. Bulgakov[3], W. Marine*[1], and M. Sentis[4]

[1] GPEC, UMR 6631 CNRS, 13288 Marseille, France.

[2] A.F. Ioffe Physico-Technical Institute, St.-Petersburg, Russia.

[3] Institute of Thermophysics, Novosibirsk, Russia.

[4] LP3, FRE 2165 CNRS, Marseille, France.

*Tel : +33 4 91 82 91 73, Fax : +33 4 91 82 91 76, E-mail : marine@gpec.univ-mrs.fr

**Abstract**

We present the results of experiments on synthesis of ZnO nanoclusters by reactive pulsed laser deposition (PLD). The nanoclusters were formed and crystallized in the gas phase and deposited on $SiO_2$ substrates. The nanostructured films were characterized by conventional photoluminescence (PL). The PL spectra consist of a narrow UV excitonic band and a broad visible band related to defects in the film. The film preparation conditions such as the substrate temperature, ambient gas nature and pressure, were optimized in order to increase the intensity of excitonic emission and prevent the formation of defects. A post-growth annealing by UV laser radiation improved the optical quality of the deposited films. The photoluminescence intensity was found to be dependent significantly on the laser fluence and on the number of shots per site. The nature of the defects responsible for the observed luminescence in a visible range is discussed.

**Introduction**

New applications in optoelectronics stimulated the research on materials for short wavelength emitting devices. ZnO is a semiconductor with a wide direct band gap (3.37 eV) and large exciton binding energy (60 meV). Exciton lasing from ZnO films at room temperature was reported recently [1]. Moreover, the efficient mirrorless laser emission with the optical feedback created by multiple light scattering (random lasing) was observed in zinc oxide microparticles [2]. To synthesize a lasing material for a random laser it is necessary to produce defect free films with strong excitonic emission, large coefficient of amplification of spontaneous emission, and strong multiple scattering [2]. The nanocrystalline ZnO films potentially possess all these features.

Recently, we showed that pulsed laser ablation and the following cluster assisted deposition is an efficient method for formation of nanoclusters and synthesis of nanocrystalline films. By varying the experimental conditions it is possible to control the film thickness, cluster sizes and their composition [3]. The main features of pulsed laser deposition are the chemical purity, low substrate temperatures, and a possibility to improve the stoichiometry by introducing a reactive atmosphere.

In this paper we present the optical properties of laser deposited ZnO films and characterize the films by conventional photoluminescence (PL). The PL spectra consist of a narrow exitonic band and a large defect-related band. We have optimized the preparation conditions and annealed the films by UV laser in order to increase the intensity of the excitonic band and decrease the defect-related emission. The visible luminescence band resulted from the transitions between the deep levels corresponding to the different local configurations of oxygen in the films [4, 5].

**Experimental**

The sintered pure ZnO target was placed on a rotating holder inside a stainless steel vacuum chamber evacuated by a turbomolecular pump. A continuous flux of oxygen mixed with helium at variable ratios was introduced into the chamber as an ambient after pumping the chamber down to about $2\times10^{-7}$ mbar. The substrate was placed in front of the target on a holder equipped with a heater, allowing variation of substrate temperature in the range of 20 – 500 °C. The ablation was performed by a pulsed ArF* laser ($\lambda$=193 nm, pulse duration 15 ns, FWHM) at a fluence level of 3.5 J/cm$^2$. The laser beam was focused onto the target with an incident angle of 45°. The PL of deposited films was excited by a Hg lamp with the wavelength of 254 nm and detected by a cooled photomultiplier linked to a grating monochromator.

We have utilized the same pulsed laser for laser annealing as for the film deposition. The laser beam was shaped by apertures and focused on the film surface at normal incidence into a rectangular spot of 18 mm$^2$.

**Results and discussion**

The nanocluster size distributions were studied by using Atomic Force Microscopy (AFM) in a tapping mode. The quantity of clusters inferior to one monolayer was deposited on an atomically plane surface of highly ordered pyrolytic graphite (HOPG). We concluded from high resolution transmission electron microscopy observations that most of the nanoparticles have a near spherical shape. The nanoparticle lateral size could not be determined from the AFM measurements due to the convolution of an object and the AFM tip shapes. Contrarily, the height of the object gives the correct size if a flat surface is used as a reference. Fig. 1 shows the size histogram of ZnO particles prepared on cleaved HOPG surface. We have presented the size statistics of about 150 nanoclusters deposited on an area of about 10 μm². The size distribution is narrow and can be described by Gaussian function with a maximum at 10 nm and a half width of 3 nm.

The sample quality has been characterized by the absolute luminescence intensity and by the ratio of excitonic to defects band intensities. The substrate temperature and the oxygen pressure were first optimized in order to prepare the films of the best quality. The quality of samples was found to be improved with the increase of the substrate temperature. However, to prevent the coalescence of clusters and their recrystallisation on the substrate surface, the temperature of 385°C has appeared to be optimal. The films prepared in 4 mbars of oxygen show the best quality. If the pressure is lower, the oxygen amount is insufficient to improve the films stoichiometry. Higher pressures strongly reduce the deposition rate and affect considerably the cluster growth process that leads to the formation of defects.

In Fig. 2 we present PL spectra of ZnO nanocrystalline films prepared on $SiO_2$ substrates heated up to 385°C. Helium partial pressure was the only variable parameter in this series of experiments. Fig. 2a shows a spectrum of the film prepared in pure oxygen without adding helium. This spectrum consists of two principal bands: a narrow UV peak of ZnO free excitons centered at photon energy of 3.28 eV with 0.1 eV FWHM, and a large visible band with maximum intensity at 2.0 - 2.4 eV related to deep defect levels.

Adding 1 mbar of He during the film preparation led to significant decrease of intensity of the defect-related band (Fig. 2b). We could also notice the increase of the excitonic to defect bands ratio of intensity, i.e. an improvement of the film optical quality. The position and the width of the excitonic peak remain the same as for the films prepared in pure oxygen. The best optical quality (strong exciton luminescence combined with a negligible defect band intensity) was achieved in a gas mixture with the partial helium pressure of 1.5 mbar (Fig. 2c). We believe that the observed PL behavior is due to a more efficient cooling of the gas phase nanoclusters through collisions with He molecules. Further augmentation of the He pressure up to 2 mbars led to decrease of absolute luminescence intensity (Fig. 2d). This was probably related to the fact that the total pressure in the preparation chamber was fairly high and thus less amount of clusters reached the substrate.

The obtained optimal ablation conditions (the fluence 3.5 J/cm², the gas mixture 4 mbar $O_2$, 1.5 mbar He) were used to produce films without heating the substrate. Typical PL spectra are presented in Fig. 3. For the as-deposited film, the exciton band is very weak and only defect-related band with a maximum at about 2 eV is observed (Fig. 3a). To improve the quality of the films we have performed annealing by the excimer laser. The PL spectra obtained after annealing the films by 30 laser shots with various fluences are presented in Figs. 3b-d.

It can be seen that annealing led to the increase of the exciton band intensity but did not affect neither its spectral position nor its shape. Contrarily, the position of the defect-related band shifted to the blue side of the spectrum.

Various mechanisms have been proposed for the orange and the green luminescence of ZnO. The orange band centered at 2 eV was observed in ZnO films with local excess of oxygen [5], and the green band at 2.4 eV is a well-known in ZnO and associated with oxygen vacancies [4, 5]. It should be noted that the maximum used fluence of 140 mJ/cm², corresponds to melting of ZnO surface. The annealing with fluences above the melting threshold changed the film morphology radically and resulted in coalescence of nanoclusters to a size of 40-60 nm [6]. The results presented in Fig. 3b and 3c correspond to annealing in the solid phase at sub-threshold fluences. Even for relatively low fluences, the augmentation of the number of laser shots reduced the intensity of the orange band and increased the green one. The excess of oxygen incorporated on the surface of clusters during film preparation can be easily removed by laser. It explains the decrease in intensity of the orange band. At the same time, the photons of 6.4 eV energy can break chemical bonds and create oxygen vacancies. This leads to increase in intensity of the green band. Once the green luminescent band obtained, it remained very stable. The increase in the number of shots or in the laser fluence did not affect the spectral position but only increased the intensity of this band.

**Conclusion**

We have developed the cluster-assisted pulsed laser deposition method in a mixed $O_2$-He background gas atmosphere to form the nanocrystalline ZnO films. The nanoclusters are formed in the gas phase and have a narrow size distribution. The films prepared with the optimized substrate temperature and gas partial pressures have an intense UV excitonic luminescence and a weak defect-related band. We have demonstrated that the optical quality of films prepared with reduced substrate temperature can be improved by UV laser annealing. We have shown possible ways to control the defect-related luminescence by changing the local amount of oxygen in the films.

Figure captions:

Fig 1. Size histogram of ZnO particles prepared on HOPG surface. The laser fluence was 3.5 J/cm²; the gas mixture was 4 mbar $O_2$ and 1.5 mbar He. A solid line is a fit with Gaussian distribution with the mean cluster size of 10 nm and dispersion of 3 nm.

Fig 2. The photoluminescence spectra of ZnO nanocrystalline films prepared on $SiO_2$ substrate at temperature of 385°C. The laser fluence was 3.5 J/cm². The oxygen and helium partial pressures were 4 mbar and: a) 0 mbar; b) 1 mbar; c) 1.5 mbar; and d) 2 mbar, respectively.

Fig 3. The photoluminescence spectra of ZnO nanocrystalline films prepared on $SiO_2$ substrate at room temperature. The laser fluence was 3.5 mJ/cm². The oxygen and helium partial pressures were 4 mbar and 1.5 mbar respectively; a) as-prepared film; b) annealed by 30 shots of ArF laser with the fluence F = 0.075 J/cm² ; c) F = 0.094 J/cm²; d) F = 0.14 J/cm². The spectra were shifted vertically for sake of clearness.

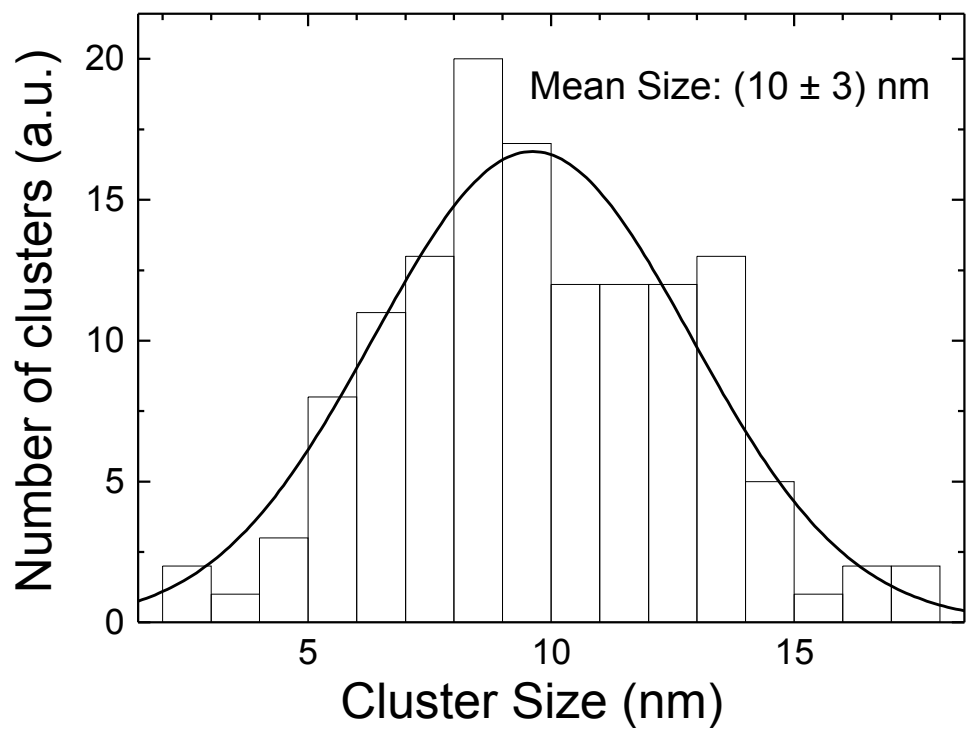

Fig 1.

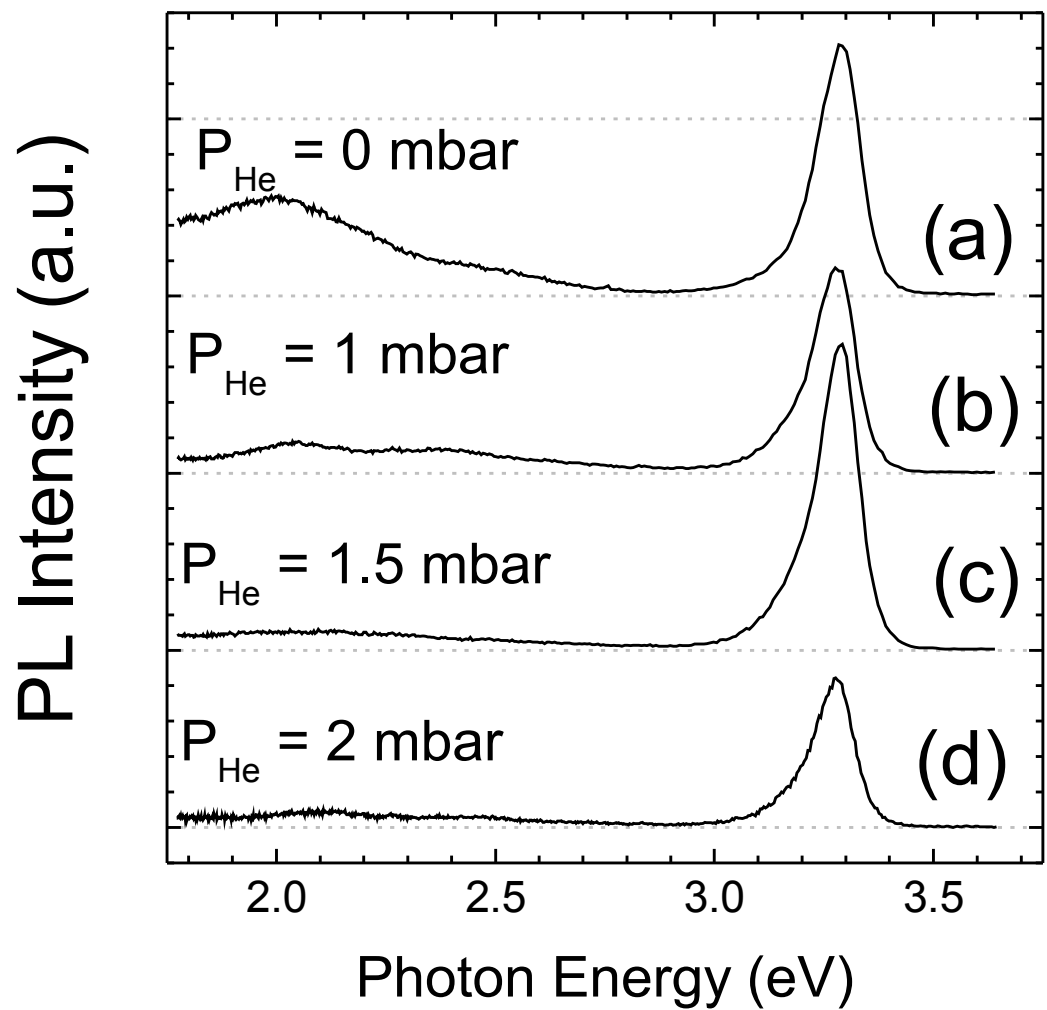

Fig 2.

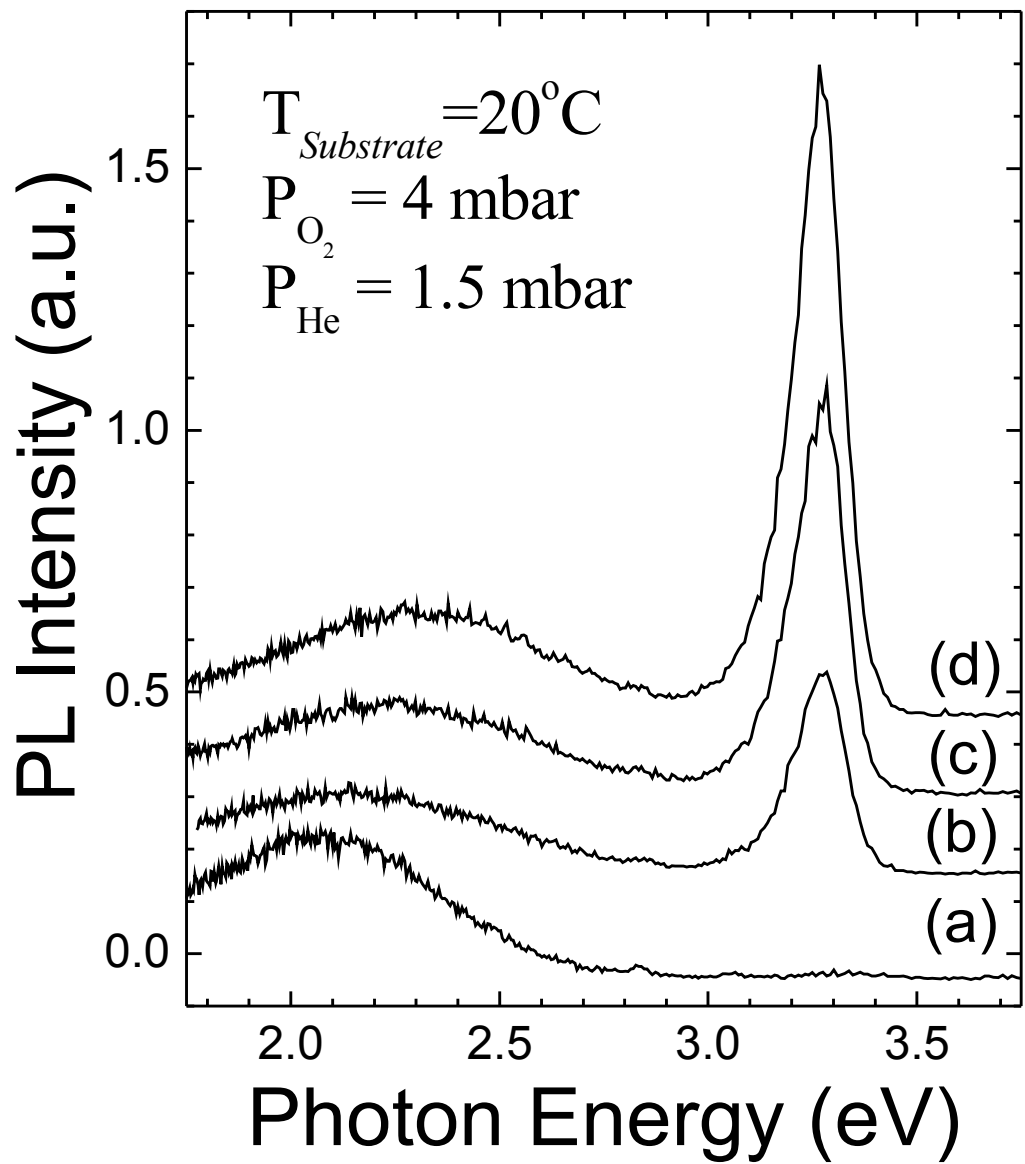

Fig 3.